\documentclass[reqno]{amsart}
\usepackage{color}
\usepackage{graphicx}
\usepackage{amsthm}
\usepackage{amsmath}
\usepackage{amssymb}
\usepackage{latexsym}
\usepackage[all]{xy}
\usepackage{tikz}
\usetikzlibrary{patterns}

\title[Investigating Causal Relationships in Stock
Returns]{Investigating Causal Relationships in Stock Returns with Temporal Logic Based Methods}
\author{Samantha Kleinberg \and Petter Kolm \and Bud Mishra}
\begin{document}
\theoremstyle{definition}
\newtheorem{defn}{Definition}[section]
\newtheorem{conj}{Conjecture}[section]
\newtheorem{exmp}{Example}[section]

\theoremstyle{plain}
\newtheorem{prop}{Proposition}[section]
\newtheorem{thm}{Theorem}[section]
\newtheorem{lem}[thm]{Lemma}
\newtheorem*{cor}{Corollary}
\newtheorem*{clm}{Claim}

\theoremstyle{remark}
\newtheorem*{rem}{Remark}
\newtheorem*{note}{Note}
\newtheorem{case}{Case}

\maketitle

\section{Causal inference in finance}

\subsection{Finance and Information}
Financial and informational systems are closely linked
through a complex network of trades and exchanges. According to the efficient market hypothesis (EMH), it seems there should be no possibility of predicting future prices by analyzing price
information from the past (weak EMH), publicly
available new information (semi-strong EMH), or
combined public and private information (strong EMH). However, rich evidence in the academic literature\footnote{See~\cite{de1985does,campbell1988stock,jegadeesh1993returns,campbell1998vrl}. For a summary see also \cite{granger1992forecasting}.} casts doubt against the unpredictability of the market. The evidence hints that there may be complex (and largely unexplored) dynamics through which information is aggregated in markets and affects price processes. Since no fully risk-free trading strategy can exploit this to produce consistent excess returns, though, this does not destabilize theories built on the EMH.

Since the resulting strategies are not risk free, we must determine the strength of predictions. The standard approach is to look at movements of
the resulting return series (expected returns, standard deviations, etc.) and their relationships (Sharpe ratio, Information
ratio, etc.). Another approach is to examine the causal relationships between predictors and their targets. To do this we need a sound and robust notion of causality. Classical causal and econometric techniques focusing on time series data typically limit inferences to pairwise relationships with a single lag between cause and effect. However, it is more likely that there is a window of time between cause and effect and further that there is a set (or even sequence) of conditions that together produce the effect.

In this paper, we discuss a new
algorithmic framework to better infer these causal relationships and
apply the method to inference of causal relationships in return time series.
In the system described, causal relationships are represented as logical formulas, which allow us to test arbitrarily complex hypotheses in a computationally efficient way.
The approach described here will allow us to combine price data with qualitative information at varying time scales, from interest rate announcements, to earnings reports to news stories and even tweets.
By integrating this information with price and volume data, we can shed light on some of the previously invisible common causes of seemingly correlated price movements.

The outline of the article is as follows. In section~\ref{sec:granger} we begin by reviewing Granger causality, a widely used method for causal inference in financial time series. In section~\ref{sec:method} we describe a new approach to causal inference, built on temporal logic and model checking. In section~\ref{sec:simulation} we discuss the development of synthetic financial time series data and in section~\ref{sec:results} we discuss the comparison of our approach to Granger causality on the generated data. Finally, we apply the methods developed to real daily stock returns.

\subsection{Granger causality}\label{sec:granger}
The primary method for inferring causality in financial applications was
developed by Granger to takes two time series and determine whether
one predicts, or causes, the other. Here,
pairwise causality is defined by~\cite{granger1980tc}:
\begin{quote}
With $\Omega_t$ being all available (non-redundant) knowledge at time
$t$, $Y_t$ \emph{Granger causes} $X_{t+1}$ if $P(X_{t+1} \in A |
\Omega_t) \neq P(X_{t+1} \in A | \Omega_t - Y_t)$ where $A$ is some
set of observations.
\end{quote}
That is, $Y_t$ provides information about
$X_{t+1}$ that is not contained in the rest of the set. There is no mention of the magnitude of the probability or how much of a
difference $Y_t$ makes to $X_{t+1}$ (there may be better
predictors or information that may
be added to $Y_t$ to improve its predictive value).
Further, there is no intrinsic method of representing complex
factors such that their causal roles may be inferred automatically
from the data.

In practice Granger causality is frequently tested using linear
regression and determining whether the use of the information in the
possible cause leads to a smaller variance in the error term than
when this information is omitted~\cite{hamilton1994}. An extension, proposed by Chen et
al.~\cite{chen2004amn} allows analysis of an arbitrary number of time
series as well as nonlinear models.
However, this definition for causality has some challenges. There is no natural method of representing a window of time between the cause and effect (versus a single lag) and while variables may be defined in an arbitrary way, it is not so simple to automatically determine their probabilities when they become complex.

\section{A new method for inference}\label{sec:method}
In contrast to prior methods, our approach to inferring causality in time series represents relationships using logical formulas (allowing explicit description of the time between cause and effect and the probability associated with the relationship) and measures causal significance by computing the average difference a cause makes to its effect given other possible causes of the same effect~\cite{kleinberg_uai09}.

\begin{figure}[h]
\caption{Example of propositions satisfying formula~\eqref{eq:form-example}.}
\label{fig:timeline}
\begin{center}
\begin{tikzpicture}
\draw[fill=gray] (0.2,1.6) rectangle (2.0,1.2);
\draw[fill=gray] (0.6,1.0) rectangle (2.2,0.6);
\draw[fill=gray] (1.5,0.4) rectangle (1.9,0);
\draw[pattern=north west lines] (4,0.4) rectangle (5.5,0);
\foreach \x in {1.5, 4, 5.5}
   \draw (\x cm,-0.4) -- (\x cm,-0.2);

\draw[->] (0,-0.3) -- (6,-0.3);
\draw (-.5,0) node[below=3pt] {\footnotesize$ time $} ;
\draw (0,1.4) node {$ a $};
\draw (0.4,0.8) node {$ b $};
\draw (1.3,0.2) node {$ c $};
\draw (1.51,-0.21) node[below=3pt] {\footnotesize$ t $} ;
\draw (3.8,0.2) node {$ d $};
\draw (4.1,-0.21) node[below=3pt] {\footnotesize$ t+t_1 $};
\draw (5.6,-0.21) node[below=3pt] {\footnotesize$ t+t_2 $};
\end{tikzpicture}
\end{center}
\end{figure}
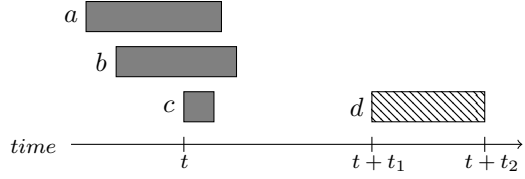

First, we represent causal relationships in a propositional probabilistic branching time temporal logic called PCTL~\cite{hansson94logic}. This logic allows us to write such formulas as
\begin{equation}\label{eq:form-example}
(a \wedge b) U c \leadsto^{\geq t_1, \leq t_2}_{\geq p} d,
\end{equation}
which could mean that after a bear market ($a$) and
increasing unemployment ($b$) persist \emph{until} unemployment reaches 20\%
($c$), then within 1 ($t_1$) to 2 ($t_2$) months, there will be a bull market
($d$) with probability $p$. Figure~\ref{fig:timeline} illustrates this formula, where solid grey bars indicate times when a proposition is true and
the patterned bar indicates the range of times when $d$ could be true and
satisfy this formula.

We could also look at the impact of world news and economic factors on particular stocks. The important feature of this type of formula is that we can systematically describe any state (such as information or other criteria) leading to (using the leads-to symbol, $\leadsto$) another state within a window of time. While methods for testing such formulas generally begin with a model, which we will not have, we have extended the approach to check such formulas over sequences of time-indexed observations (called traces). These sequences could include the daily returns for hundreds of stocks, company announcements such as earnings reports, and federal announcements such as interest rate changes.

These formulas will be our causal hypotheses. We may generate all possible relationships between, say, pairs of stocks, or between all of the basic variables in our system, up to some specified formula size.
Then, we determine which of these are satisfied by our data and are potential, or \emph{prima facie}, causes.
\begin{defn}\label{def:my-pf}$c$ is a \emph{prima facie} cause of $e$ if the following conditions all hold:
\begin{enumerate}
  \item $F_{>0}^{\leq \infty} c$,
  \item $c\leadsto^{\geq 1,\leq \infty}_{\geq p} e$, and
  \item $F_{<p}^{\leq \infty} e$.
\end{enumerate}
\end{defn}
We have used the PCTL $F$ (finally) and $\leadsto$ (leads-to) operators, which are defined as follows. First, $F_{\geq p}^{\leq t} g$ means that $g$ will eventually be true within $t$ time units with probability $p$. Then, with $f$ and $g$ being logical formulas, $f \leadsto_{\geq p}^{\geq r, \leq s} g$ means that after $f$ is true, $g$ will be true in between $r$ and $s$ time units, with probability $p$.
Our conditions mean that some $c$ causes some $e$ (where both are logical formulas) if $c$ is earlier than $e$ by at least one time unit, $c$ has a non-zero probability, and the probability of $e$ given $c$ is greater than the unconditional probability of $e$. However, finding something that only occurs earlier and raises the probability of the effect admits many false causes. Thus, we need to further test these potential causes, assessing the relative significance of each.

We compare each cause, pairwise, with all other possible cause of the same effect as follows. With $X$ being the set of all other \emph{prima facie} causes of $e$, we compute
\begin{equation}\label{eq:eps-avg}
    \epsilon_{avg}(c,e)=\frac{\displaystyle\sum_{x \in X \setminus c} \epsilon_x(c,e)}{|X\setminus c|},
\end{equation}
where
\begin{equation}\label{eq:eps}
\epsilon_x(c,e)=P(e|c\wedge x)-P(e|\neg c \wedge x).\end{equation}
This means that for each \emph{prima facie} cause of each effect, we average the difference it makes to the probability of its effect given each other \emph{prima facie} cause of that effect. Then, we can partition causes into significant and insignificant classes based on their values for  $\epsilon_{avg}$.
\begin{defn}\label{def:spur-tl}
A prima facie cause, $c$, of an effect, $e$, is an
$\epsilon$-\emph{insignificant cause} of $e$ if
$|\epsilon_{avg}(c,e)|< \epsilon$.
\end{defn}

\begin{defn}\label{def:gen-tl}
A prima facie cause, $c$, of an effect, $e$, that is not an
$\epsilon$-\emph{insignificant cause} of $e$ is an $\epsilon$-\emph{significant}, or \emph{just-so}, cause.
\end{defn}

Note that a cause that is seemingly significant (large value of $\epsilon_{avg}$) may not in fact be the genuine cause of its effect. If we have data only on the returns of stocks, but not the underlying market conditions causing the movements of stock prices, we might only find proxies for these true causes. That is, if some stocks appear to react to, say, interest rate changes earlier than other stocks do, and these changes affect all stocks, those that respond earlier will appear to cause the price movements of those that respond later. While they do not truly cause these movements, they can however be used for predictive purposes.

We must now determine an appropriate value for $\epsilon$. While we could potentially use prior knowledge of the problem or simulations to find this, another approach is to treat this as a multiple hypothesis testing problem, where we aim to control our false discovery rate (FDR)\cite{benjamini1995cfd}. While most approaches to FDR control require prior assumptions and knowledge in order to construct a null hypothesis, we can use the large amount of hypotheses tested to infer this null distribution from the data itself.
We begin by assuming that if the data contain no true causal relationships, then the computed $\epsilon_{avg}$ values will follow a normal distribution due to the large number of hypotheses tested. When there are true causal relationships, these will be governed by another distribution and will appear as deviations from the underlying normal. Thus we can recognize the significant relationships by finding those that differ substantially from a normal after fitting such a distribution to the observed data.

More formally, we follow the empirical Bayesian formulation introduced by Efron \cite{efron2004lss}, which may be summarized as follows. We begin with $N$ hypotheses $H_1, H_2 \ldots H_N$ and their corresponding $z$-values $z_1, z_2 \ldots z_N$. These scores are the number of standard deviations by which a test result deviates from the mean, so that with mean $\mu$ and standard deviation $\sigma$, the $z$-value for a result $x$ is simply $(x-\mu)/\sigma$. In analyzing our results, the $z$-values will be derived from the computed $\epsilon_{avg}$'s. Then, the results fall into two classes, null (corresponding to causes that are spurious or where the effects are too small to be of interest) and non-null (corresponding to causes that are genuine or large enough to be interesting), where we respectively accept and reject the null hypothesis. We assume the proportion of non-null cases is small relative to $N$, say no more than $10\%$. Note that if this assumption fails, which can happen in practical applications, we cannot reliably estimate the null empirically. That is because the mixture of the two classes will be significantly skewed by the non-null class and will no longer be likely to follow a normal distribution.

The prior probabilities of a case being in the null or non-null class are
 $p_0$ and $p_1=1-p_0$, with the densities of each class ($f_0(z)$ and $f_1(z)$) describing the
distribution of these probabilities. When using a theoretical null, $f_0(z)$ is
the standard $N(0,1)$ density. We define the mixture density
\begin{equation}
f(z)=p_0f_0(z) + p_1f_1(z), \label{eq:mixture}
\end{equation}
then the posterior probability of a case being uninteresting given $z$ is
\begin{equation}
Pr\{null|z\} = p_0f_0(z)/f(z), \label{eq:posterior-int}
\end{equation}
and the \emph{local false discovery rate}, is
\begin{equation}
fdr(z)\equiv f_0(z)/f(z). \label{eq:local-fdr}
\end{equation}
Note that, in this formulation, the $p_0$ factor is ignored, yielding an upper
bound on $fdr(z)$. Assuming that $p_0$ is large (close to 1), this
simplification does not lead to massive overestimation of $fdr(z)$. The entire procedure is then:
\begin{enumerate}
  \item Estimate $f(z)$ from the observed $z$-values;
  \item Define the null density $f_0(z)$ either from the data or
  using the theoretical null;
  \item Calculate $fdr(z)$ using equation \eqref{eq:local-fdr};
  \item Label $H_i$ where $fdr(z_i)$ is less than a threshold (say,
  0.01) as interesting, or in our case, causally significant.
\end{enumerate}

\section{Simulation of financial time series}\label{sec:simulation}
To compare the proposed inference approach to existing work, we developed a set of simulated financial time series, allowing us to embed a variety of causal relationships in the data and see how well each algorithm is able to recover these.
For this purpose, we used a common factor model that allowed two kinds of causality: one through the influence of common factors on stock portfolios and the other a direct dependency between individual portfolios. Our simulated market consisted of 25 portfolios, with data generated for six scenarios during two different 3001 day time periods. In each scenario factors could be shifted (in time) for each individual portfolio and there could be dependency between portfolios. We initially assume that a portfolio's return at time $t$ depends on the values of the factors at time $t-3$, making it possible to test whether factors may be treated as common causes of portfolio returns.

The six portfolios, summarized in table~\ref{tbl:data}, contain three (A-C) with no dependency between individual portfolios, and three (D-F) where three such relationships were included. Then, each portfolio in the set can have its factors lagged the same amount (A,D), half the portfolios may be lagged by a different amount (B,E) or half the portfolios may be lagged by a random amount in the range [0,3] lags, where each factor for a portfolio can be lagged independently of the others (C,F).
\begin{table}
\begin{centering}
\begin{tabular}{c|r r r }
  Name & One lag & Random lag & Portfolio dependency \\
  \hline
  A &  & & \\
  B &  \checkmark &  & \\
  C &  &\checkmark & \\
  D & & &\checkmark \\
  E & \checkmark& & \checkmark\\
  F & &\checkmark & \checkmark\\
\end{tabular}
\caption{Summary of datasets created. Half the portfolios in a dataset may have their factors lagged by a single amount (one lag), or half may have each individual factor lagged by a random amount in $[0,3]$ (random lag). When dependency between portfolios is included, there are three portfolios whose return at $t_i$ depends on the returns of another portfolio at $t$.}
\label{tbl:data}
\end{centering}
\end{table}
The return of portfolio $i$ at time $t$ is then given by
\begin{equation}\label{eq:return}
r_{i,t}=\sum_j \beta_{ij}f_{j,t'} + \epsilon_{i,t},
\end{equation}
where factor $j$ at time $t$ is denoted $f_{j,t}$. In case $A$, $t'=t-3$. In
cases D, E, and F, $\epsilon$ is the sum of the randomly generated error plus,
in the case where portfolio $i$ depends on portfolio $k$, $\epsilon_{k,t-1}$.
To construct these simulated series, we used the Fama-French three factor model~\cite{fama1993common}, and the $5 \times 5$ size/book-to-market
portfolios~\cite{fama:web}; both using the daily data series.
Specifically, we regressed the 25 book-to-market portfolios onto the market, HML, and SMB factors (see~\cite{fama1993common} for the definition of these factors; the factors can be downloaded from~\cite{fama:web}) and estimated the empirical distribution of the regression coefficients $\beta_{ij}$ and correlations of the resulting residuals of the portfolios by bootstrapping over different time periods. Our simulated return series data of the scenarios A through F was then generated by randomly drawing betas and residual correlations from these empirical distributions, and then by applying~\eqref{eq:return} for two non-intersecting time periods of 3001 daily observations the market, HML, and SMB factors.

\section{Empirical results and discussion}\label{sec:results}
\subsection{Data and method}
We compared the algorithm of Kleinberg and Mishra (called \verb"AITIA") with the MSBVAR \verb"granger.test" function in R. In order to assess the algorithms as well as some common assumptions and practices, we conducted a series of tests on the twelve datasets described above: one using the generated returns (sequences of the returns $r_{i,t}$ as defined in \eqref{eq:return}), one using the actual error terms used to construct the returns (sequences of residuals $\epsilon_{i,t}$ as defined in \eqref{eq:return}), one using the generated returns with the known factors (sequences of factors $f_{j,t}$) included to give a total of 28 variables (25 portfolios plus three factors), and finally one comprised of residuals calculated by regressing the returns on the known factors (approximating a common approach to such time series).
For both algorithms we tested pairwise relationships between elements of the time series (portfolios, and in some cases factors) at lags of 1, 2, and 3 days. For \verb"AITIA", this meant testing whether a positive/negative return for one variable caused a positive/negative return in another. Since the Granger implementation only returned the significance of a relationship between variables (regardless of whether it was positive or negative), true positives were broadly defined as being that there is a causal relationship between two variables in a certain amount of time.

The procedure for each was to define the set of causal relationships to be tested and then run each algorithm to compute the significance of each relationship in this set, resulting in a set of $\epsilon_{avg}$'s for \verb"AITIA" and $F$-statistics with their associated $p$-values for \verb"granger.test". Then, the empirical null hypotheses and false discovery rates for each test were computed and relationships with an fdr$<0.01$ called significant. For \verb"AITIA" we used the \verb"locfdr" R package from Efron~\cite{efron2008locfdr} to compute the null hypothesis, while we found that due to the different distribution, the \verb"fdrtool" package~\cite{strimmer2008fdrtool} provided better results for \verb"granger.test".

\subsection{Results}
In order to compute FDR and FNR rates, we must understand what constitutes a true positive. In the simplest case, when using the generated returns, there should be no causal relationships found in scenario A, while in B and C we should find that portfolios with lags less than $t-3$ should cause those with greater lags, with the time associated with the relationship being that of the difference between the lags. In datasets D-F, our findings should be the same, with the addition of the embedded relationships between portfolios. While the way the data is generated may make it seem that the factors could cause the portfolio returns, examination of the factors reveals that this is not the case. Recall that the Fama-French factors are constructed from the stocks themselves, thus when we lag the factors, this can be interpreted as if some portfolios respond to external influences and affect the market factors earlier than others.

In the datasets consisting only of the error terms, we should find only the embedded relationships between portfolios (since there is no influence from factors in these time series). Similarly, when we look at the residuals, we expect that the result should ideally be the same as that for the error terms and no influence from factors should remain. However, in practice, due to estimation errors, the returns are not so cleanly split into factor/error terms and the result of regressing on the factors and removing this component is not the same as the original error terms. We confirmed that this is the case, and the relationships between lagged and unlagged portfolios persist. This data was not used for computation of error rates. Finally, we can also include the factors in the dataset used for inferring causal relationships. If the factors were not derived from the stocks, we would find them to be common causes of the lagged portfolios. However, since they may be viewed as both cause and effect, it becomes difficult to determine what would constitute a true positive.

For our assessment of the two algorithms we focus on the return data (which in addition to being the most straightforward was also the one on which both algorithms performed best). We will briefly discuss the idiosyncratic term data, but for the reasons mentioned above, the residual and combined portfolio/factor experiments do not lend themselves to rigorous quantitative assessment.
\begin{table}
\begin{centering}
\begin{tabular}{c|r r r }
  Method & FDR & FNR & Intersection \\
  \hline
  \verb"AITIA" & 0.0775 & 0.0417 & 0.8090 \\
  Granger & 0.6547 & 0.0863 & 0.4347 \\
\end{tabular}
\caption{Comparison of results on synthetic financial data.}
\label{tbl:comparison-fin}
\end{centering}
\end{table}

Results including FDR, FNR and intersection for the generated returns are shown in table~\ref{tbl:comparison-fin}.
Note that the FDR for \verb"AITIA" is an order of magnitude lower than that for the Granger test.
These values are across all twelve datasets (two for each scenario), and include relationships with all levels of lags. We also compare how consistent our results are by computing the intersection of relationships found in both time ranges for a particular scenario. Since the only causal relationships in the system are those we embed, the relationships found should be the same. Using \verb"AITIA" the intersection was nearly 81\% while for granger.test it was just over 43\%. On the idiosyncratic (or error) returns, the FDRs were quite high, owing to the fact that there are extremely few true positives (0 in A-C, and 3 in each of D-F) and potentially some subtle dependencies between portfolios. The FDRs were high for both algorithms (0.827 for \verb"AITIA" and 0.988 for Granger), though the 18 true positives were found by both, with both having zero false negatives. We found a substantial difference in the quantities of false discoveries. While the rates were high for both algorithms, \verb"AITIA" made 86 false discoveries (out of 104 total) while Granger made 1442 (out of 1460).

\begin{figure}
  \includegraphics[width=4in]{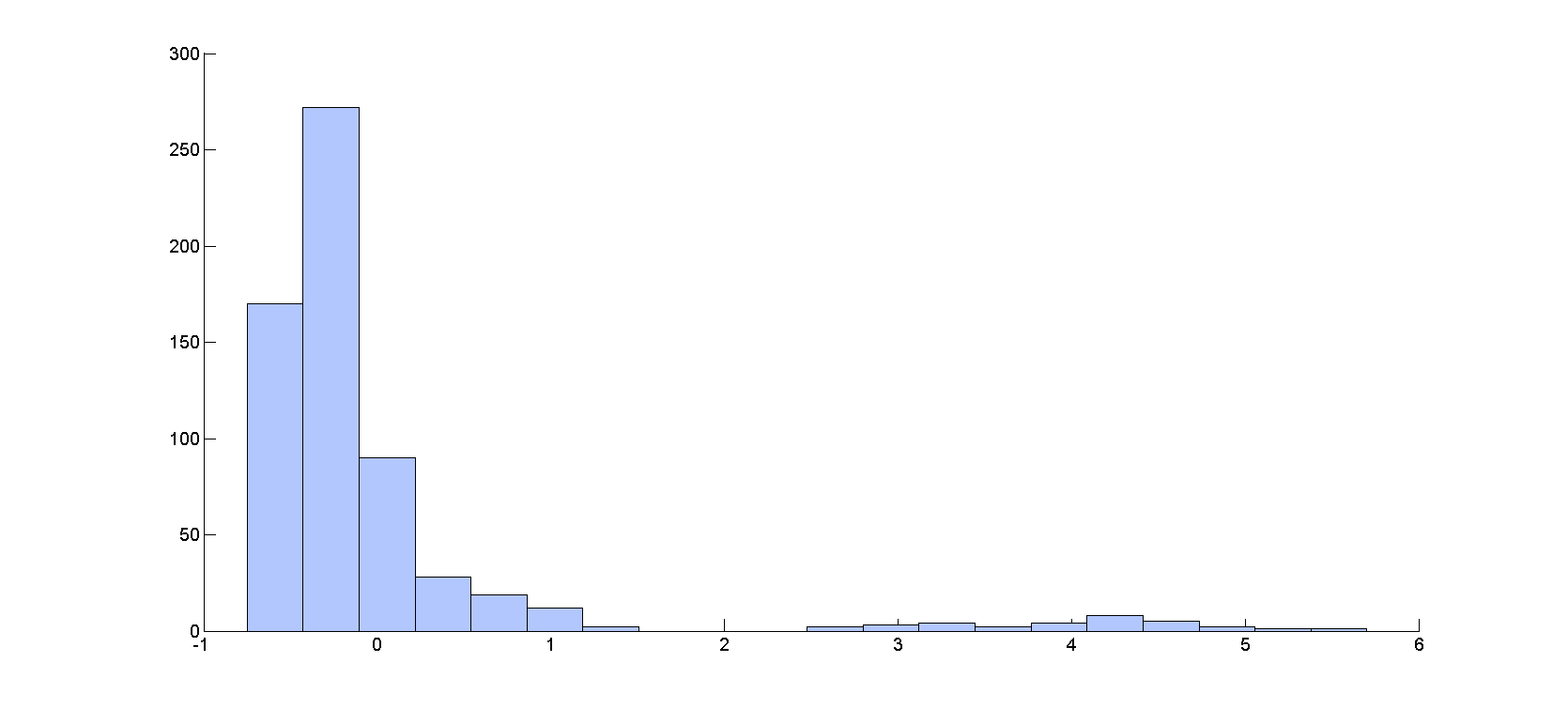}
  \caption{Histogram of $z$-values (computed from the set of $\epsilon_{avg}$ values for one test.}\label{fig:C13}
\end{figure}

We note that while our FDR on the returns data is substantially lower than that of \verb"granger.test", it is still higher than our desired rate of 0.01. This is due to difficulties in correctly inferring the null distribution. Since the number of true positives can be substantial (in some cases much greater than the 10\% frequently assumed), we violate one of the assumptions of these methods: that our observation is mostly from the null distribution and that there are a small number of deviations from that, corresponding to non-nulls. In fact in many cases visual inspection of the graphs reveals that a human could clearly see the separation between the two classes. For an example, see figure~\ref{fig:C13}. When we allow for manual choice of thresholds, our FDR is reduced below our specified target, with a negligible (0.6\%) increase in false negatives. The results also become quite consistent (intersection of greater than 98\%), meaning that the true positives are found in both time ranges (and that it is possible to improve results by calling significant only those causes found significant in both). Further work on empirical null methods will be necessary to bring automated analysis closer to this ideal.

\subsection{Real data}
To determine how similar actual market data is to our synthetic
returns, we tested our algorithm on daily stock returns using the
CRSP database, downloaded through WRDS. We began with all stocks in the S\&P 500 for the entirety of January 1, 2000 to
December 31, 2007, yielding over 2000 trading days. We tested random subsets of 100 stocks in this
set. Over the entire time period, we found no significant
relationships (using fdr$<0.01$) when testing for pairwise
relationships between stocks at a timescale of one day. Looking at
the last 800 trading days, we found a single significant relationship,
and again found zero looking at the last 400 trading days. One
explanation for the few discoveries made is that at the timescale of
one day and over long periods of time, relationships between
companies do not persist (and are overshadowed by market-wide
factors).

\begin{figure}
  \includegraphics[width=5in]{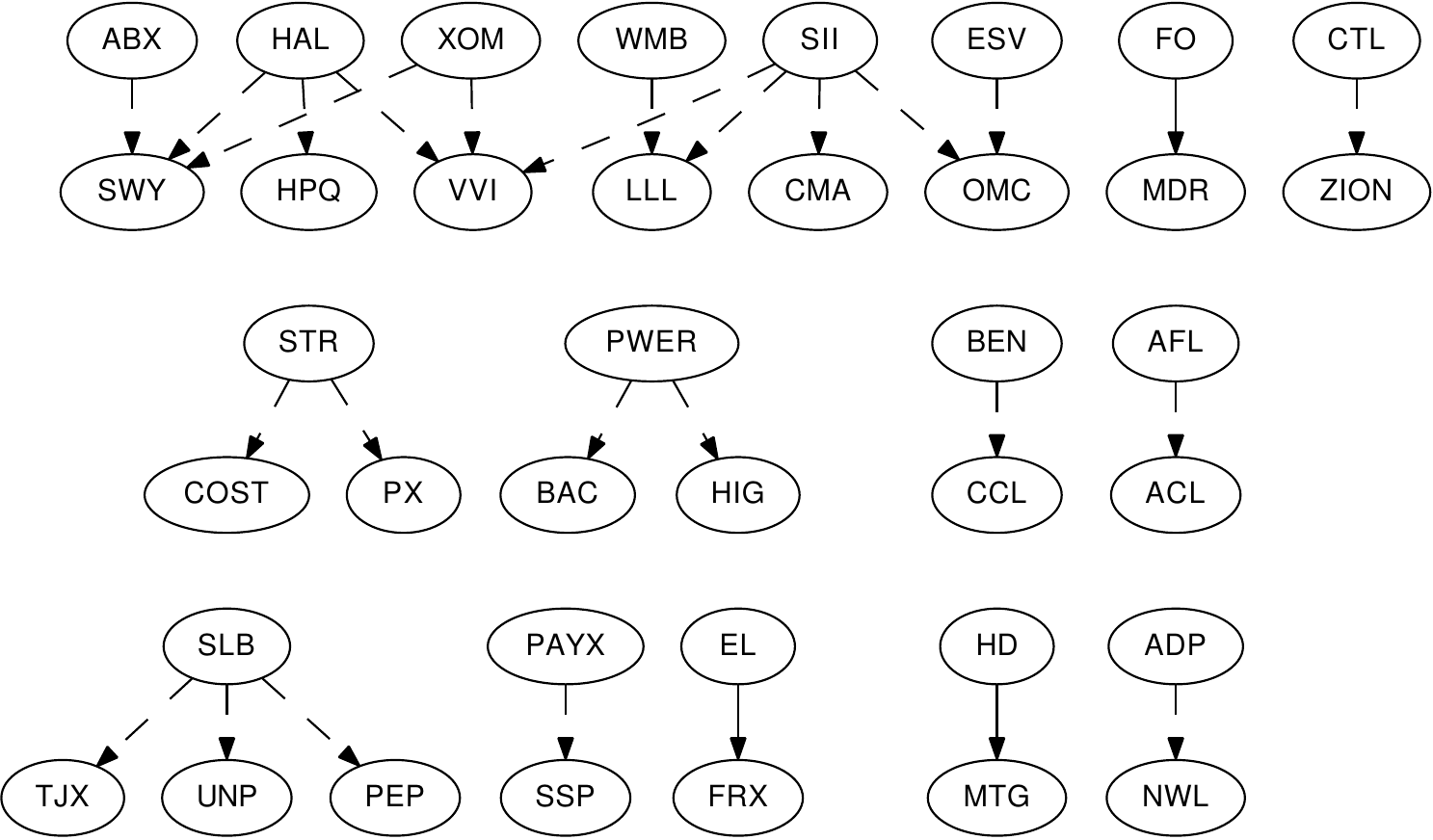}
  \caption{Relationships found in one year of real daily market returns, January 1, 2007 through December 31, 2007. A dashed arrow from $x$ to $y$ denotes that ``a price increase in $x$ causes a price
decline in $y$ in exactly 1 day'', while a solid arrow from $x$ to $y$ denotes ``a price increase in $x$ causes a
price increase in $y$ in exactly 1 day''.
  }\label{fig:real-rels}
\end{figure}

Finally, we focused on one year of trading, using the last 252
trading days from the series. Due to the shorter time series, we
examined a larger set of stocks: those that were in the S\&P 500
during the 2000-2007 time period. There were 386 such stocks and 27
significant relationships, which are shown
in figure~\ref{fig:real-rels}. These relationships are
primarily of the form ``a price increase in $x$ causes a price
decline in $y$ in exactly 1 day'' (denoted by a dashed line in the
figure) consistent with one-day reversals (see~\cite{lo1990contrarian,jegadeesh1995short}), with a few of the form ``a price increase in $x$ causes a
price increase in $y$ in exactly 1 day'' (denoted by a solid line in
the figure). Many of the causes in this set are companies involved
in oil, gas and energy, while financial companies appear to be
influenced by returns of stocks from the technology sector.

\section{Conclusions}
Understanding and accurately inferring causality is vital in finance, where we
aim to understand the relationships between stocks and predict how the market
will behave.
We have shown that by staying close to
philosophical theories of causality, translating these into the framework of
temporal logic and model checking and applying statistical methods for false
discovery control, our method remains computationally feasible while
significantly outperforming traditional approaches that reflect only
correlation.
While the construction of optimal trading rules from inferred causal
relationships remains to be determined, there are a few probable approaches.
First, it is likely that causal relationships may improve standard predictive regressions and pairs trading
strategies. Second, the relationships can be used to
aid portfolio construction  and risk management, since the causal relationships between possible components
will be known. For instance, there might be situations where multiple causally related stocks are included in a portfolio but none of the standard risk factors are able to pick this up.

\bibliography{references}
\bibliographystyle{abbrv}
\end{document}